\documentclass{article}
\usepackage{geometry}
\usepackage{amsmath}
\usepackage{graphicx}

\title{Anomalous stochastic transport of particles with self-reinforcement and Mittag-Leffler distributed rest times}
\author{Daniel Han \footnote{Department of Mathematics, University of Manchester, Oxford Rd, Manchester, M13 9PL, UK;} , Dmitri V. Alexandrov \footnote{Department of Theoretical and Mathematical Physics, Ural Federal University, Lenin Ave., 51, Ekaterinburg, 620000, Russian Federation;} , Anna Gavrilova  \footnote{School of Biological Sciences, University of Manchester, Manchester, M13 9PL, UK}\\ and Sergei Fedotov $^{*}$}

\begin{document}
	\maketitle
	\abstract{We introduce a persistent random walk model for the stochastic transport of particles involving self-reinforcement and a rest state with Mittag-Leffler distributed residence times. The model involves a system of hyperbolic partial differential equations with a non-local switching term described by the Riemann-Liouville derivative. From Monte Carlo simulations, this model generates superdiffusion at intermediate times but reverts to subdiffusion in the long time asymptotic limit. To confirm this result, we derive the equation for the second moment and find that it is subdiffusive in the long time limit. Analyses of two simpler models are also included, which demonstrate the dominance of the Mittag-Leffler rest state leading to subdiffusion. The observation that transient superdiffusion occurs in an eventually subdiffusive system is a useful feature for application in stochastic biological transport.}
	
	\section{Introduction}
	
	The stochastic movement of intracellular organelles, cells and animals very often exhibits anomalous diffusion, which has led to the widespread use of fractional diffusion equations and fractional derivatives in modelling \cite{metzler2000random,mendez2010reaction}. There are several recent observations that emphasize the importance of fractional models in biological phenomena, such as cancer cell motility \cite{huda2018levy}, polarized cell dynamics \cite{estrada2021motility}, intracellular transport of organelles \cite{fedotov2018memory} and animal migration \cite{reynolds2018current}. In particular, we observe superdiffusive and subdiffusive transport simultaneously in intracellular transport \cite{kenwright2012first,fedotov2018memory,han2020deciphering}. Recently, the superdiffusion was modelled by a persistent random walk using the concept of self-reinforcing directionality \cite{chen2015memoryless,han2021self}. However, it is well known that endosomes often rest before moving and that these rest times are power-law distributed \cite{han2020deciphering}. Therefore, it is natural to formulate the self-reinforcing persistent random walk model with Mittag-Leffler distributed rest times, which have power-law tails.
	
	For continuous time random walks (CTRW), the competition between power-law run and rest times has been explored thourougly \cite{portillo2011intermittent,zaburdaev2015levy,klafter2011first}. A single model based on the elephant random walk \cite{schutz2004elephants} with reinforcement exhibiting superdiffusion, diffusion and subdiffusion in the long time limit has been formulated in discrete time and space \cite{kumar2010memory}. The model presented here is actually a generalization of the elephant random walk \cite{schutz2004elephants,da2014ultraslow,boyer2014solvable,baur2016elephant,bercu2019hypergeometric,bercu2019multi,da2020non}, a jump process, to a persistent random walk framework with finite velocity \cite{goldstein1951diffusion,mendez2010reaction,rossetto2018one}. Using the persistent random walk framework is advantageous as extensions such as reactions, chemosensitive movement and interactions between agents are established in literature \cite{othmer1988models,hillen2002hyperbolic,fort2002wavefronts,filbet2005derivation,fetecau2010investigation,bouin2014hyperbolic,perthame2016derivation,calvez2019chemotactic,kumar2021multiscale} and convenient to introduce. The purpose of our paper is to explore the impact of an anomalous rest state on self-reinforced persistent random walks with finite velocity.
	
	\section{Stochastic transport with self-reinforcement and Mittag-Leffler distributed rest times}
	
\begin{figure}
		\centering
		\includegraphics[width=0.6\linewidth]{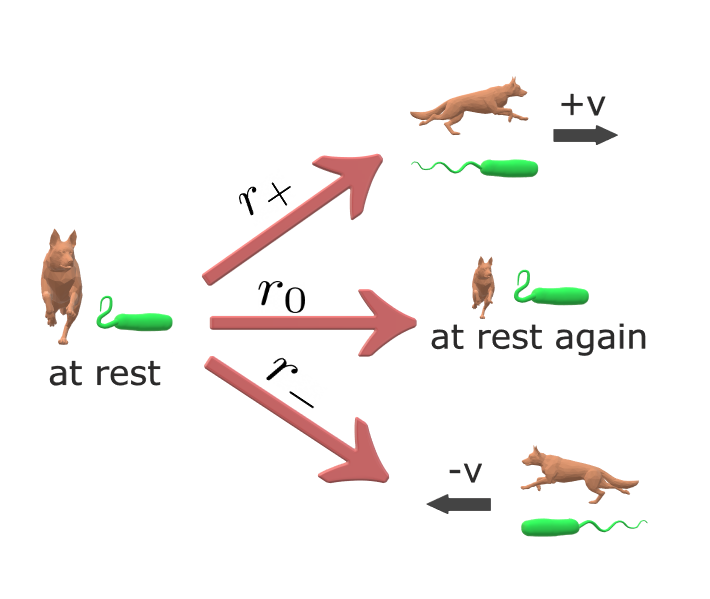}
		\caption{A diagram that shows a dog and a monotrichous bacterium making the decision to transition from the rest state to either the positive velocity, rest and negative velocity states with the associated conditional transition probabilities, $r_+$, $r_-$ and $r_0$. If the conditional transition probabilities depend on the previous history of choices, $r_+$, $r_-$ and $r_0$ should depend on the current position and age.}
		\label{fig:diagram_transitions}
	\end{figure}

	To implement rests to the self-reinforcing persistent random walk \cite{han2021self}, we  formulate a model with three states. We introduce the probability density functions (PDFs) for the active states with positive and negative velocity, $p_+(x,t)$ and $p_-(x,t)$, and the resting state, $p_0(x,t)$. In the active states, the random walk runs with constant speed $\nu$ for an exponentially distributed time with rate $\lambda$. After each active run, the random walk pauses for a Mittag-Leffler distributed residence time and then makes a choice to switch to some next state. With conditional transition probability $r_+$, $r_-$ and $r_0$, the random walk transitions from rest to the positive velocity state, negative velocity state or remains in the rest state ($r_++r_-+r_0=1$). An illustration of this is shown in Figure \ref{fig:diagram_transitions}. The PDEs that represent this random walk are
	\begin{equation}
		\begin{split}
			\frac{\partial p_{\pm}}{\partial t} & \pm \nu \frac{\partial p_{\pm}}{\partial x} = -\lambda p_{\pm} + r_{\pm} i(x,t), \\
			\frac{\partial p_0}{\partial t} & = \lambda p_+ +\lambda p_- - (1-r_0) i(x,t),
		\end{split}
		\label{pdesystem_powerlawrests}
	\end{equation}
	where the integral escape rate from the rest state, $i(x,t)$, is defined as follows (see for example, \cite{angstmann2013continuous,fedotov2013nonlinear,angstmann2021general})
	\begin{equation}
		i(x,t) = \tau_0^{-\beta}\mathcal{D}_t^{1-\beta} p_0(x,t).
		\label{integralescaperate_definition}
	\end{equation}
	Here, the Riemann-Liouville derivative is
	\begin{equation}
		\mathcal{D}_t^{1-\beta}p_0(x,t)=\frac{1}{\Gamma(\beta)} \frac{\partial}{\partial t}\int_{0}^{t} \frac{p_0(x,t')}{(t-t')^{1-\beta}}dt' \hspace{0.3cm}\text{  (}0<\beta<1\text{)}.
	\end{equation}
	Note that for the non-Markovian alternating states, one can use the general expression for the escape rate  $i(x,t)$ in the form of convolution of the memory kernel and density
	\cite{fedotov2011non}.
	
	In what follows, we will introduce and explain the two key components in the model \eqref{pdesystem_powerlawrests}: self-reinforcement and Mittag-Leffler distributed rest times.
	
	\textit{Self-reinforcement.} Firstly, introducing self-reinforcement to \eqref{pdesystem_powerlawrests} requires careful consideration of the conditional transition probabilities. 
	What really is self-reinforcement and how does one introduce it in \eqref{pdesystem_powerlawrests}? If a process is self-reinforcing, then it retains a memory of its past decisions and uses this history to affect its current decisions. In other words, the random walk should consider its past decisions to transition into positive and negative velocity states and adjust the conditional transition probabilities ($r_+$, $r_-$ and $r_0$). Looking at Figure \ref{fig:diagram_transitions}, the particle or animal should have $r_+$, $r_-$ and $r_0$ change with time as more decisions are made. 
	
	An intuitive formulation of self-reinforcement is that the time spent travelling with velocity $\pm\nu$ determines the probability that the particle will switch to that state. Mathematically,
	\begin{equation}
		r_{\pm} = w_1 \frac{t_{\pm}}{t} + w_2 \frac{t_{\mp}}{t}+w_3\frac{t_0}{t} \hspace{0.3cm} \text{and} \hspace{0.3cm} r_0 = w_3,
		\label{conditionaltransitionprobabilities}
	\end{equation}
	where $t_+/t$, $t_-/t$ and $t_0/t$ are the relative times spent in the positive, negative and zero velocity states respectively, out of the total time elapsed $t$. The constant prefactors $w_1$, $w_2$ and $w_3$ are weights on each of those relative times and $w_1+w_2+w_3=1$.
	To avoid the trivial case when the reinforcement to the rest state results in permanent rest, we set $w_3=1/3$. Now using $t = t_++t_-+t_0$ and $x=x_0+\nu(t_+ -t_-)$ in \eqref{conditionaltransitionprobabilities}, $r_{\pm}$ can be expressed as
	\begin{equation}
		r_{\pm} = \frac{1}{3}\pm\alpha_0\frac{x-x_0}{2\nu t} \hspace{0.3cm} (\alpha_0 = w_1-w_2),
		\label{conditionaltransitionprobabilities2}
	\end{equation}
	where $\alpha_0$ the self-reinforcement parameter. The initial position of the random walk is $x_0$ but for simplicity, we will assume $x_0=0$ from now on. If $\alpha_0 >0$, then $w_1 >w_2$ and time spent in corresponding states increases probabilities of future occupation in that state (in \eqref{conditionaltransitionprobabilities} $r_{\pm}$ increases more as $t_{\pm}$ increases since $w_1>w_2$). On the other hand if $\alpha_0<0$, then $w_1<w_2$ and time spent in the corresponding states increases probabilities of future avoidance of that state ($r_\pm$ increases more as $t_{\mp}$ increases since $w_1<w_2$). Intuitively, reinforcement of past behaviour can be represented by $w_1>w_2$ and punishment of past behaviour as $w_1<w_2$. The equation \eqref{conditionaltransitionprobabilities2} is a powerful formulation since we can express self-reinforcement as an additive term to constant and equal conditional transition probabilities of $1/3$. Moreover, this additive term encapsulates the past history of the random walk by accounting for the ratio of current position $x$ and current time $t$. This ratio compares how far the particle has moved away from the initial position given the maximum possible position it could have obtained. It is now clear that the superdiffusion generated by this self-reinforcing mechanism \cite{han2021self} is fundamentally different to those generated by power-law flights in CTRW and L\'evy walk formulations (a biological motivation is given in Section VII of \cite{han2021self}). It would be interesting to extend this model to three dimensions as it was done for the diffusion-advection model for intracellular transport in \cite{lin2021modelling}.
	
	\textit{Mittag-Leffler distributed rest times.} Secondly, power-law distributed rest times with the divergent first moment are introduced into \eqref{pdesystem_powerlawrests} through the Riemann-Liouville fractional derivative. This method  is well established in literature \cite{mendez2010reaction}. This component is significant as power-laws are seen in many empirical observations for stochastic processes which possess complex underlying mechanisms \cite{clauset2009power}. Pertinent examples of power-laws include the waiting times between: stock transactions \cite{sabatelli2002waiting}, arrivals of internet viruses \cite{liebovitch2003information}, sudden decreases in terminal airway resistance for lungs with respiratory problems \cite{suki1994avalanches}, players joining a game network \cite{henderson2001modelling}, household residence before moving \cite{fedotov2017anomalous}; consecutive emails sent \cite{barabasi2005origin}; dopamine signalling in \textit{Drosophila melanogaster} \cite{ueno2012dopamine}; and active-passive state switching in endosome movement \cite{han2020deciphering}. All of these examples, demonstrate the importance of power-law waiting times in real phenomena, highlighting the need for a self-reinforcing persistent random walk model with an agent that takes power-law distributed rests while deciding the choice of the next run. To be exact, the introduction of this integral escape rate, $i(x,t)$, is equivalent to the random walker waiting in the rest state for a random time, which has the probability density
	\begin{equation}
		\psi(\tau) = -\frac{d}{d\tau}E_{\beta}\left[-\left(\frac{\tau}{\tau_0}\right)^{\beta}\right] \hspace{0.3cm} \text{where}\hspace{0.3cm}0<\beta<1,
		\label{MLfunction}
	\end{equation}
	and $E_{\beta}(\cdot)$ is the one parameter Mittag-Leffler function. The waiting times in the rest state will be distributed approximately like $(\tau/\tau_0)^{-\beta}$ for large values of $\tau/\tau_0$ resulting in `heavy' or power-law tails.
	
	Now, from \eqref{pdesystem_powerlawrests}, we can formulate a single governing equation by introducing the total density function $p=p_++p_-+p_0$ and the flux $J=\nu (p_+ - p_-)$. Then
	\begin{equation}
		\begin{split}
			\frac{\partial p}{\partial t} & =- \frac{\partial J}{\partial x}, \hspace{0.5cm}
			\frac{\partial J}{\partial t} = - \nu^2 \frac{\partial p}{\partial x} + \nu^2\frac{\partial p_0}{\partial x} -\lambda J + \alpha_0\frac{x}{t} i(x,t),\\
			\frac{\partial p_0}{\partial t} & = \lambda p-\lambda p_0  - (1-r_0) i(x,t).
		\end{split}
		\label{pJidentities}
	\end{equation}
	We can eliminate the flux $J$ by combining the first two equations in \eqref{pJidentities} and arrive at the governing equations
	\begin{equation}
		\begin{split}
			\frac{\partial^2 p}{\partial t^2} &+\lambda \frac{\partial p}{\partial t}  = \nu^2 \frac{\partial ^2 p}{\partial x^2}-\nu^2\frac{\partial^2 p_0}{\partial x^2}-\frac{ \alpha_0 }{ t} \frac{\partial}{\partial x}\left[x \tau_0^{-\beta}\mathcal{D}_t^{1-\beta} p_0(x,t)\right],\\
			\frac{\partial p_0}{\partial t} &= \lambda p - \lambda p_0 - (1-r_0)\tau_0^{-\beta}\mathcal{D}_t^{1-\beta} p_0(x,t),
		\end{split}
		\label{governingequations}
	\end{equation}
	where for self-reinforcement, $0<\alpha_0<2/3$ and $r_0=1/3$. From these governing equations, it is not immediately clear what the effect of Mittag-Leffler distributed rest times will have when competing with self-reinforcement. To elucidate this relationship, we perform second moment calculations in the next section.
	
	\section{Second moment calculations}
	
\begin{figure}
		\centering
		\includegraphics[width=10.5cm]{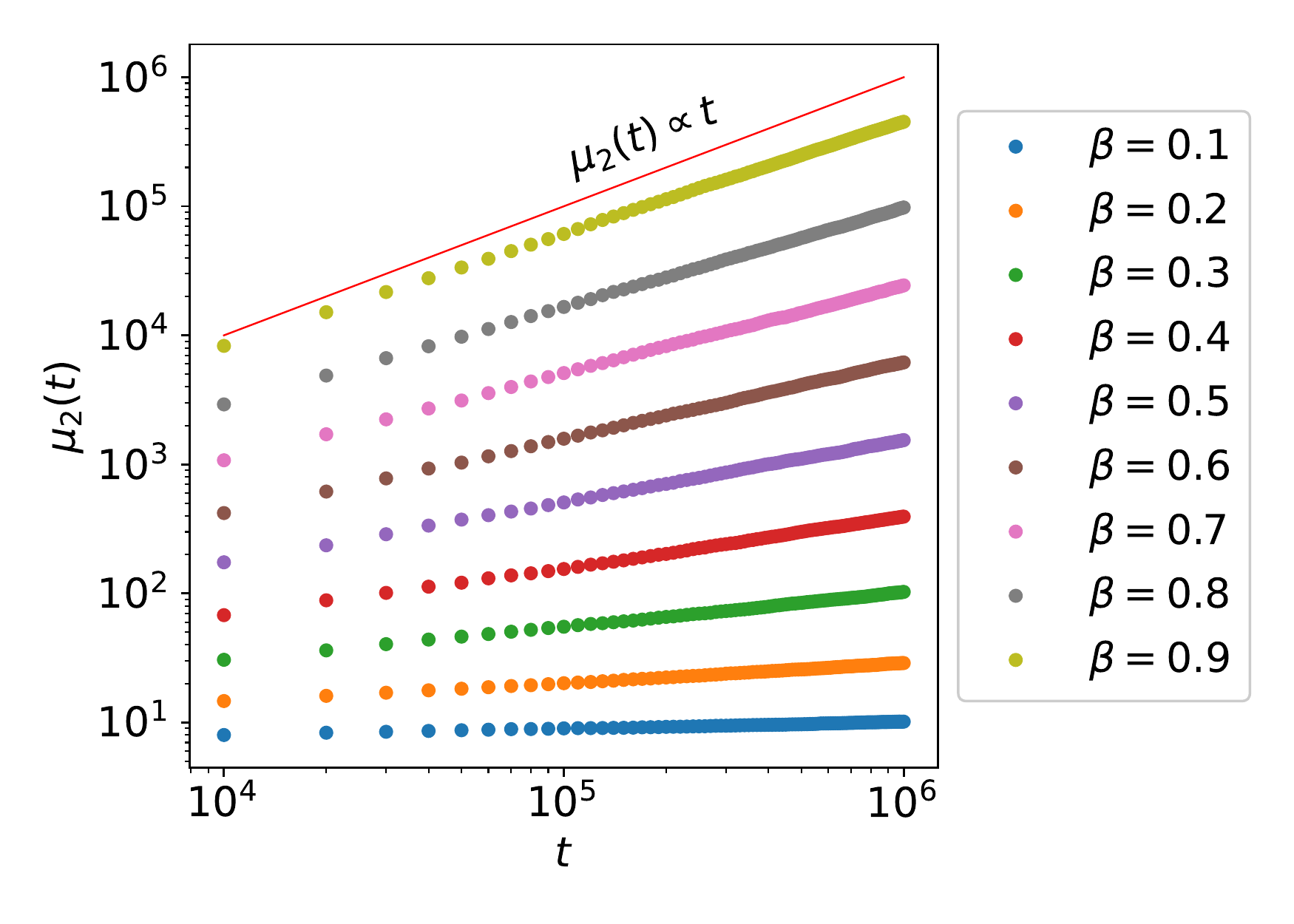}
		\caption{Plots of second moments for Monte Carlo simulated trajectories. Each value of $\beta$ has $N=10^4$ trajectories. The parameters used were $\alpha_0 = 0.6$, $\nu=1$, $\lambda =1$ and $\tau_0=1$. }
		\label{fig:1}
	\end{figure}

	From \eqref{governingequations}, we obtain the fractional differential equations
	\begin{equation}
		\begin{split}
			\frac{d^2 \mu_2}{dt^2}& + \lambda \frac{d\mu_2}{dt} = 2\nu^2\left[1-N_0\right] + \frac{2\alpha_0}{t}\tau_0^{-\beta}\mathcal{D}_t^{1-\beta}\mu_{20},\\
			\frac{d\mu_{20}}{dt}& = \lambda\mu_2-\lambda\mu_{20}- (1-r_0)\tau_0^{-\beta}\mathcal{D}_t^{1-\beta}\mu_{20},
		\end{split}
		\label{secondmomentpde}
	\end{equation}
	where $\mu_2=\int_{-\infty}^{\infty}x^2pdx$, $\mu_{20}= \int_{-\infty}^{\infty}x^2 p_0dx$ and $N_0 = \int_{-\infty}^{\infty}p_0dx$. For $N_0$, we can obtain from \eqref{governingequations} \begin{equation}
		\frac{dN_0}{dt} = \lambda -\lambda N_0 - (1-r_0)\tau_0^{-\beta}\mathcal{D}_t^{\beta}N_0.
		\label{N0ODE}
	\end{equation}
	Equation \eqref{N0ODE} is a fractional differential equation describing the total probability to find the random walk in the rest state.
	To simplify calculations, we take the Laplace transform of \eqref{secondmomentpde} and \eqref{N0ODE} along with the initial conditions $p_+(x,0)=\delta(x)$, $p_-(x,0) = 0$ and $p_0(x,0)=0$ to obtain
	\begin{equation}
		\begin{split}
			-(s^2+&\lambda s)\frac{d\hat{\mu}_2}{ds} - (2s+\lambda)\hat{\mu}_2=\frac{2\nu^2}{s^2} + 2\nu^2 \frac{d\hat{N}_0}{ds} + 2\alpha_0\tau_0^{-\beta}s^{1-\beta}\hat{\mu}_{20},\\
			s\hat{\mu}_{20} &= \lambda \hat{\mu}_2 -\lambda \hat{\mu}_{20} - (1-r_0)\tau_0^{-\beta}s^{1-\beta}\hat{\mu}_{20},\\
			s\hat{N}_0& = \frac{\lambda}{s}-\lambda\hat{N}_0 - (1-r_0)\tau_0^{-\beta}s^{1-\beta}\hat{N}_0.
		\end{split}
		\label{secondmomentpdeLaplace}
	\end{equation}
	Rearranging and taking the long time limit, $t\rightarrow\infty$ ($s\rightarrow0$), the equations in \eqref{secondmomentpdeLaplace} give
	\begin{equation}
		\begin{split}
			-\lambda s\frac{d\hat{\mu}_2}{ds}&-\lambda \hat{\mu}_2 \approx \frac{2\nu^2}{s^2} + 2\nu^2\frac{d\hat{N}_0}{ds} + 2\alpha_0 \tau_0^{-\beta}s^{1-\beta}\hat{\mu}_{20},\\
			\hat{\mu}_{20} \approx \hat{\mu}_2 &\left(1- \frac{1-r_0}{\lambda}\tau_0^{-\beta}s^{1-\beta}\right), \hspace{0.3cm}
			\hat{N}_0 \approx s^{-1}\left(1-\frac{1-r_0}{\lambda}\tau_0^{-\beta}s^{1-\beta}\right).
		\end{split}
		\label{secondmomentpdeLaplaceasymptotic}
	\end{equation}
	Using \eqref{secondmomentpdeLaplaceasymptotic} we obtain a single equation for the second moment in Laplace space
	\begin{equation}
		-s\frac{d\hat{\mu}_2}{ds} - \hat{\mu}_2 \approx \frac{2\nu^2(1-r_0)\beta}{\lambda^2}\tau_0^{-\beta}s^{-1-\beta}.
		\label{secondmoment_laplacefinalode}
	\end{equation}
	From \eqref{secondmoment_laplacefinalode}, we obtain the asymptotic second moment in Laplace space as
	\begin{equation}
		\hat{\mu}_2(s) \sim \frac{2\nu^2(1-r_0)}{\lambda^2}\tau_0^{-\beta}s^{-1-\beta}, \hspace{0.3cm} s\rightarrow 0.
		\label{secondmoment_laplaceasymptotic}
	\end{equation}
	Finally, taking the inverse Laplace transform, 
	\begin{equation}
		\mu_2(t) \sim \frac{2\nu^2 (1-r_0)}{\lambda^2\Gamma(1+\beta)}\left(\frac{t}{\tau_0}\right)^{\beta}, \hspace{0.3cm} t\rightarrow \infty.
		\label{secondmomentasymptotic}
	\end{equation}
	Clearly, \eqref{secondmomentasymptotic} demonstrates that the random walk with self-reinforcement \eqref{conditionaltransitionprobabilities2} and Mittag-Leffler distributed rest times \eqref{MLfunction} is subdiffusive in the long time limit. This theoretical result is confirmed by the second moment of numerical simulations shown in Figure \ref{fig:1}. Interestingly, transient superdiffusion is found in Monte Carlo simulations shown in Figure \ref{fig:3}. This suggests that self-reinforcement still plays a major role at shorter time scales but is negated by the eventual trapping of particles in the rest state. This type of behaviour is important in intracellular transport, where organelles transition between superdiffusion and subdiffusion at different time scales \cite{korabel2018non,han2020deciphering}.
	
	To understand intuitively what occurs when introducing a heavy tailed waiting time for the rest state, we consider two simple cases: a single velocity model and a symmetric two velocity model both with a non-Markovian rest state, and derive their second moments in the long time limit. 
	
	\subsection{Single active state model}
	
	As a first example, consider the simplest possible case where there is only one active state with velocity $\nu$ and one rest state, such that the system of fractional differential equations describing this random walk is 
	\begin{equation}
		\begin{split}
			\frac{\partial p_+}{\partial t} &+\nu \frac{\partial p_+}{\partial x} = -\lambda p_+ +\tau_0^{-\beta}\mathcal{D}_t^{1-\beta}p_0\\
			\frac{\partial p_0}{\partial t} &= \lambda p_+ -\tau_0^{-\beta}\mathcal{D}_t^{1-\beta}p_0.
		\end{split}
		\label{toymodel1_pde}
	\end{equation}
	This simple fractional model can be used to describe the movement of intracellular organelles in only one direction interrupted by rests with Mittag-Leffler distributed residence times.
	
	From \eqref{toymodel1_pde}, we obtain a system of fractional differential equations describing the second moment
	\begin{equation}
		\frac{d\mu_{2+}}{dt} -2\nu\mu_{1+} = -\lambda\mu_{2+} +\tau_0^{-\beta}\mathcal{D}_t^{1-\beta}\mu_{20}, \hspace{0.3cm}\frac{d\mu_{20}}{dt} = \lambda\mu_{2+}-\tau_0^{-\beta}\mathcal{D}_t^{1-\beta}\mu_{20},
		\label{toymodel1_secondmoment}
	\end{equation}
	where $\mu_{2+} =\int_{-\infty}^{\infty}x^2p_+dx$, $\mu_{1+} =\int_{-\infty}^{\infty}xp_+dx$ and other symbols were defined in \eqref{secondmomentpde}. From adding together the equations in \eqref{toymodel1_secondmoment}, we find
	\begin{equation}
		\frac{d\mu_2}{dt} = 2\nu\mu_{1+},
		\label{toymodel1_secondmomentexpression}
	\end{equation}
	where $\mu_2 = \mu_{2+} + \mu_{20}$. In order to find $\mu_{1+}$, we again use \eqref{toymodel1_pde} to obtain
	\begin{equation}
		\frac{d\mu_{1+}}{dt} - \nu N_+ = -\lambda \mu_{1+} + \tau_0^{-\beta}\mathcal{D}_t^{1-\beta}\mu_{10}, \hspace{0.3cm}
		\frac{d\mu_{10}}{dt} = \lambda \mu_{1+} - \tau_0^{-\beta}\mathcal{D}_t^{1-\beta}\mu_{10},
		\label{toymodel1_firstmoment}
	\end{equation}
	where $N_+ = \int_{-\infty}^{\infty}p_+dx$ and $\mu_{10} = \int_{-\infty}^{\infty}xp_0dx$. Recall that our main objective is to find $\mu_{2}$ for which we need $\mu_{1+}$, but from \eqref{toymodel1_firstmoment} it is clear that we also need to find $N_+$. So, we integrate \eqref{toymodel1_pde} to obtain
	\begin{equation}
		\frac{dN_+}{dt}  = -\lambda N_+ + \tau_0^{-\beta}\mathcal{D}_t^{1-\beta}N_0(t), \hspace{0.3cm}
		\frac{dN_0}{dt} = \lambda N_+ - \tau_0^{-\beta}\mathcal{D}_t^{1-\beta}N_0.
		\label{toymodel1_zerothmoment}
	\end{equation}
	To derive equations \eqref{toymodel1_secondmoment} \eqref{toymodel1_firstmoment} and \eqref{toymodel1_zerothmoment}, we have used the fact that $p_+(x,t) =p_0(x,t) = 0$ and $dp_+/dx = dp_0/dx =0$ as $x\rightarrow\pm\infty$. This is because $p_+$ and $p_0$ are probability density functions that must be normalizable, and additionally, we know that this random walk propagates with finite speed from some initial position. 
	
	It is clear that dealing with the Riemann-Liouville derivative in Laplace space will be far easier than attempting to solve \eqref{toymodel1_secondmomentexpression}, \eqref{toymodel1_firstmoment} and \eqref{toymodel1_zerothmoment} directly. For initial conditions, we assume that the random walk starts in the active state at $x=0$ at $t=0$ such that $p_+(x,0) = \delta(x)$ and $p_0(x,0) = 0$. In a similar way to deriving \eqref{secondmomentpdeLaplace}, using the initial conditions in conjunction with the Laplace transforms of \eqref{toymodel1_secondmomentexpression}, \eqref{toymodel1_firstmoment} and \eqref{toymodel1_zerothmoment}, we can obtain
	
	
	\begin{equation}
		\hat{\mu}_2(s) \sim \frac{2\nu^2}{\lambda^2}\tau_0^{-2\beta}s^{-2\beta-1}.
		\label{toymodel1_secondmomentlaplaceasymptotic}
	\end{equation}
	Finally, taking the inverse Laplace transform
	\begin{equation}
		\mu_2(t) \sim \frac{2\nu^2}{\lambda^2\Gamma(2\beta+1)}\left(\frac{t}{\tau_0}\right)^{2\beta}.
	\end{equation}
	Using an analogous procedure as above, the first moment for this model can be calculated as $\mu_1(t) \sim \nu t^{\beta}/\lambda \tau_0^{\beta} \Gamma(\beta+1)$. This draws parallels with the fractional Poisson process \cite{laskin2003fractional,mainardi2007fractional,mainardi2007beyond,beghin2010poisson,cahoy2010parameter,meerschaert2011fractional,politi2011full}, which has exactly the same time dependence for the first and second moments. In fact closely examining \eqref{toymodel1_pde}, we can see the underlying stochastic process for the single active state model is the fractional Poisson process. For \eqref{toymodel1_pde}, the random walk waits in a rest state for a Mittag-Leffler distributed random time and then proceeds to travel with finite velocity $\nu$ for an exponentially distributed random time.
	
	\subsection{Bi-directional transport model}
	
	The second example we will consider is an extension of the first by adding another active state with velocity $-\nu$. This model is ideal for bi-directional intracellular transport \cite{smith2001models} with intermittent resting for power-law distributed times. The system of equations that describes this random walk is
	\begin{equation}
		\begin{split}
			\frac{\partial p_{\pm}}{\partial t} &\pm\nu \frac{\partial p_{\pm}}{\partial x} = -\lambda p_{\pm} +\frac{1}{2}\tau_0^{-\beta}\mathcal{D}_t^{1-\beta}p_0,\\
			\frac{\partial p_0}{\partial t} &= \lambda p_+ + \lambda p_- -\tau_0^{-\beta}\mathcal{D}_t^{1-\beta}p_0.
		\end{split}
		\label{toymodel2_pde}
	\end{equation}
	In order to derive an expression for the second moment, we combine the equations in \eqref{toymodel2_pde}, as we did in \eqref{pdesystem_powerlawrests} to obtain \eqref{pJidentities}. Doing this, we find
	\begin{equation}
		\begin{split}
			\frac{\partial p}{\partial t} & +  \frac{\partial J}{\partial x} = 0, \hspace{0.5cm}
			\frac{\partial J}{\partial t} + \nu^2 \frac{\partial p}{\partial x} - \nu^2\frac{\partial p_0}{\partial x} = -\lambda J,\\
			\frac{\partial p_0}{\partial t} & = \lambda p-\lambda p_0  - \tau_0^{-\beta}\mathcal{D}_t^{1-\beta}p_0.
		\end{split}
		\label{toymodel2_pJidentities}
	\end{equation}
	Again, we can eliminate $J$ by combining the first two equations in \eqref{toymodel2_pJidentities} to arrive at the governing equations
	\begin{equation}
		\begin{split}
			\frac{\partial^2 p}{\partial t^2}& +\lambda\frac{\partial p}{\partial t}= \nu^2\frac{\partial^2p}{\partial x^2}-\nu^2\frac{\partial^2 p_0}{\partial x^2},\\
			\frac{\partial p_0}{\partial t} & = \lambda p -\lambda p_0 - \tau_0^{-\beta}\mathcal{D}_t^{1-\beta}p_0.
		\end{split}
		\label{toymodel2_governingequations}
	\end{equation}
	We can obtain similar equations to \eqref{secondmomentpde} and \eqref{N0ODE} using \eqref{toymodel2_governingequations}
	\begin{equation}
		\begin{split}
			\frac{d^2 \mu_2}{dt^2} &+ \lambda \frac{d\mu_2}{dt} = 2\nu^2 - 2\nu^2 N_0, \hspace{0.5cm}
			\frac{d\mu_{20}}{dt}  = \lambda \mu_2 - \lambda \mu_{20}-\tau_0^{-\beta}\mathcal{D}_t^{1-\beta}\mu_{20},\\
			\frac{dN_0}{dt} &= \lambda - \lambda N_0 -\tau_0^{-\beta}\mathcal{D}_t^{1-\beta}N_0.
		\end{split}
		\label{toymodel2_secondmoment}
	\end{equation}
	Similar to the first example, 
	using the initial conditions $p_+(x,0)=\delta(x)/2$, $p_-(x,0)=\delta(x)/2$ and $p_0(x,0)=0$, and taking the asymptotic limit as $s\rightarrow0$, we obtain
	\begin{equation}
		\hat{N}_0(s) \approx \frac{\lambda}{s}\frac{1}{\lambda+\tau_0^{-\beta}s^{1-\beta}}, \hspace{0.5cm}
		\hat{\mu}_2(s) \approx \frac{2\nu^2}{\lambda s}\left[\frac{1}{s}-\hat{N}_0(s)\right].
		\label{toymodel2_momentprogresslaplaceasymptotic}
	\end{equation}
	After substitution, we can find the second moment for the second example in the long time limit as
	\begin{equation}
		\hat{\mu}_2(s) \sim \frac{2\nu^2}{\lambda^2} \tau_0^{-\beta} s^{-1-\beta}.
		\label{toymodel2_secondmomentlaplaceasymptotic}
	\end{equation}
	Finally, taking the inverse Laplace transform
	\begin{equation}
		\mu_2(t) \sim \frac{2\nu^2}{\lambda^2\Gamma(\beta+1)}\left(\frac{t}{\tau_0}\right)^{\beta},
		\label{toymodel2_secondmomentasymptotic}
	\end{equation}
	where $0<\beta<1$. Again, we find that the power-law rests dominate for long times and generates subdiffusion.
	
	For symmetric active states with velocities $\pm\nu$, we find that the second moment is purely subdiffusive unlike the first example. In the first example, there was no fractional diffusion limit that could be taken. However, in this second example, the fractional diffusion limit exists, which means that the system can be approximated accurately by the fractional diffusion equation
	\begin{equation}
		\frac{\partial p}{\partial t} = D_{\beta} \frac{\partial^2}{\partial x^2}\mathcal{D}_t^{1-\beta}p,
	\end{equation}
	where the fractional diffusion coefficient has an explicit form
	\begin{equation}
		D_{\beta} = \frac{\nu^2}{\lambda^2\Gamma(\beta+1)\tau_0^{\beta}}.
	\end{equation}
	Therefore in the second example, we see that the introduction of a non-Markovian rest state with divergent mean residence time causes the second moment to be subdiffusive. On the other hand, the first example is inherently different because there is no accurate fractional diffusion equation to approximate the system and so the second moment should not be interpreted in diffusive, subdiffusive or superdiffusive terms.
	
	Comparing \eqref{secondmomentasymptotic} with \eqref{toymodel2_secondmomentasymptotic}, we see that the only difference is the constant multiplier $1-r_0$. The non-Markovian rest state (or equivalently, the Mittag-Leffler distributed waiting times for rests) completely neutralizes the superdiffusion generated by the self-reinforcement in the long time limit. Figure \ref{fig:1} demonstrates this by calculating the second moment from simulated trajectories of the random walk corresponding to the system of PDEs in \eqref{governingequations}. 
	
	\section{Monte Carlo Simulations}
	
	Here, we perform Monte Carlo simulations of the random walk corresponding to \eqref{pdesystem_powerlawrests} to demonstrate that the second moment exhibits subdiffusion. The procedure for simulations is:
	\begin{enumerate}
		\item Initialize variables for current time $T_c=0$, particle position $X_c=0$ and current state $S_c=1$. The possible values for $S_c$ are $0$, $1$ or $-1$ corresponding to the rest, positive velocity and negative velocity states respectively. For convenience, we assume the random walk starts with $S_c=1$. 
		\item Set the constants: $\lambda$, $\beta$, $\tau_0$, $\nu$, $\alpha_0$, $r_0$ and $t_{end}$, the end time of simulation.
		\item If $S_c = 0$, generate a random number $\Delta T = - \lambda \ln(U) [\sin(\pi\beta)/\tan(\pi\beta V)-\cos(\pi\beta)]^{1/\beta}$, where $U,V\in[0,1)$ are uniformly distributed random numbers (see Eq. (20) in \cite{fulger2008monte}). Otherwise, generate a random number $\Delta T = -\ln(U)/\lambda$.
		\item Increment simulation time $T_c = T_c+\Delta T$ and particle position $X_c = X_c + \nu S_c \Delta T$. 
		\item If  $S_c = \pm1$, then set $S_c=0$.
		Otherwise, do the following:
		\begin{itemize}
			\item[-] if $0 \leq W < R_+ $, set $S_c=1$;
			\item[-] if $R_+\leq W<R_++R_- $, set $S_c = -1$;
			\item[-] otherwise, set $S_c=0$;
		\end{itemize}
		where $W\in[0,1)$ is a uniformly distributed random number and $R_{\pm} = r_0\pm\alpha_0 X_c/(2\nu T_c)$. 
		\item Iterate steps 3 to 5 until $T_c \geq t_{end}$.
	\end{enumerate}
	All simulations were performed using Python3. To reduce execution time, the `Numba' package and the `multiprocessing' package were used for JIT compilation and CPU parallelization, respectively.
	
\begin{figure}
		\centering
		\includegraphics[width=10.5cm]{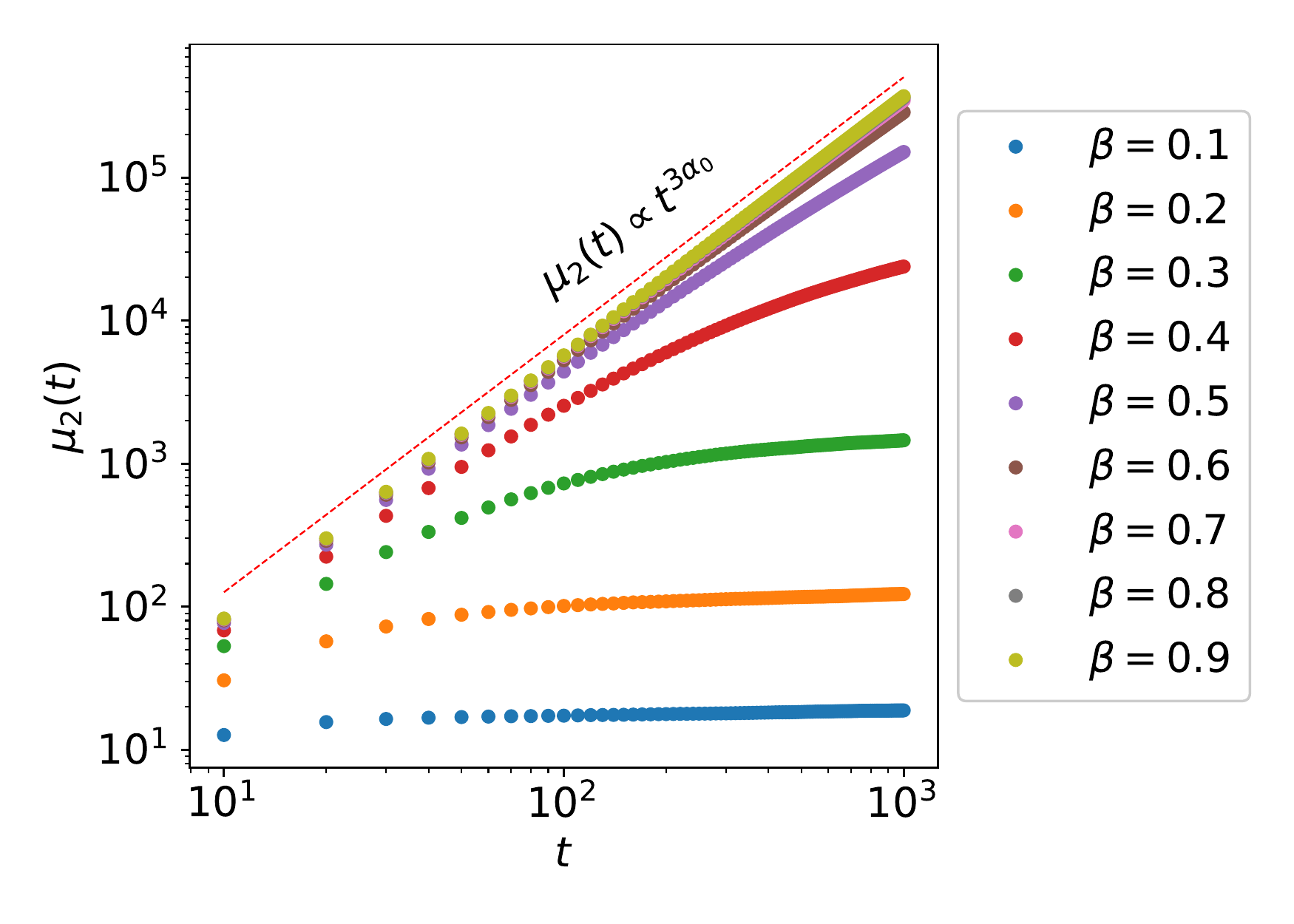}
		\caption{Plots of second moments for Monte Carlo simulated trajectories. Each value of $\beta$ has $N=10^4$ trajectories. The parameters used were $\alpha_0 = 0.6$, $\nu=1$, $\lambda =1$ and $\tau_0=10^{-4}$.}
		\label{fig:3}
	\end{figure}

	Figure \ref{fig:1} confirms the analytical result in \eqref{secondmomentasymptotic}, which predicted subdiffusion for the long time limit regardless of the strength of self-reinforcement, in this case $\alpha_0=0.6$. Interestingly, Figure \ref{fig:3} shows that if $\tau_0$ is small, then superdiffusion is possible for intermediate times.
	
\begin{figure}
		\centering
		\begin{minipage}[c]{.35\textwidth}
			\centering
			$\tau_0=1$
			\includegraphics[width=\linewidth]{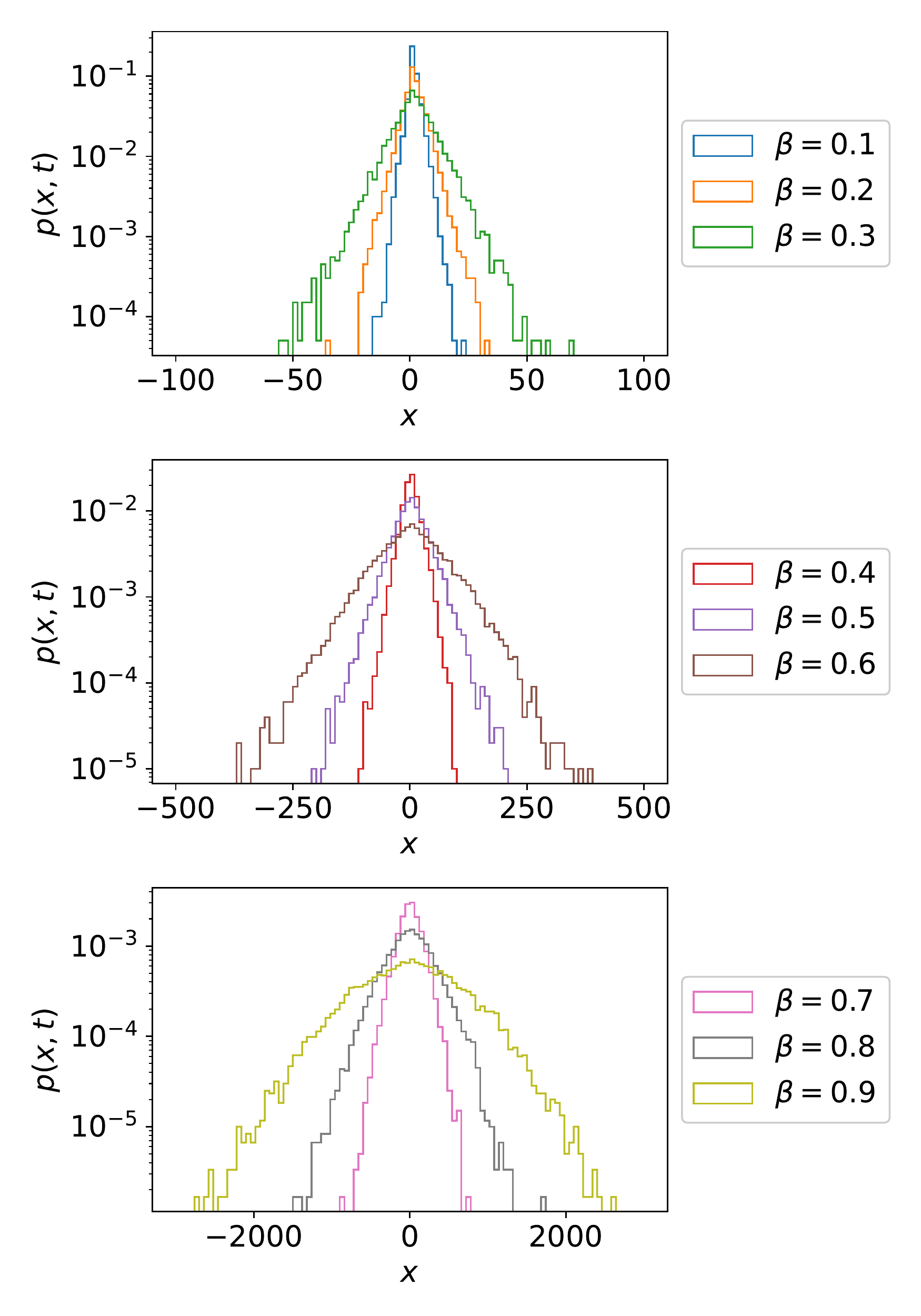}
		\end{minipage}%
		\begin{minipage}[c]{0.35\textwidth}
			\centering
			$\tau_0=10^{-4}$
			\includegraphics[width=\linewidth]{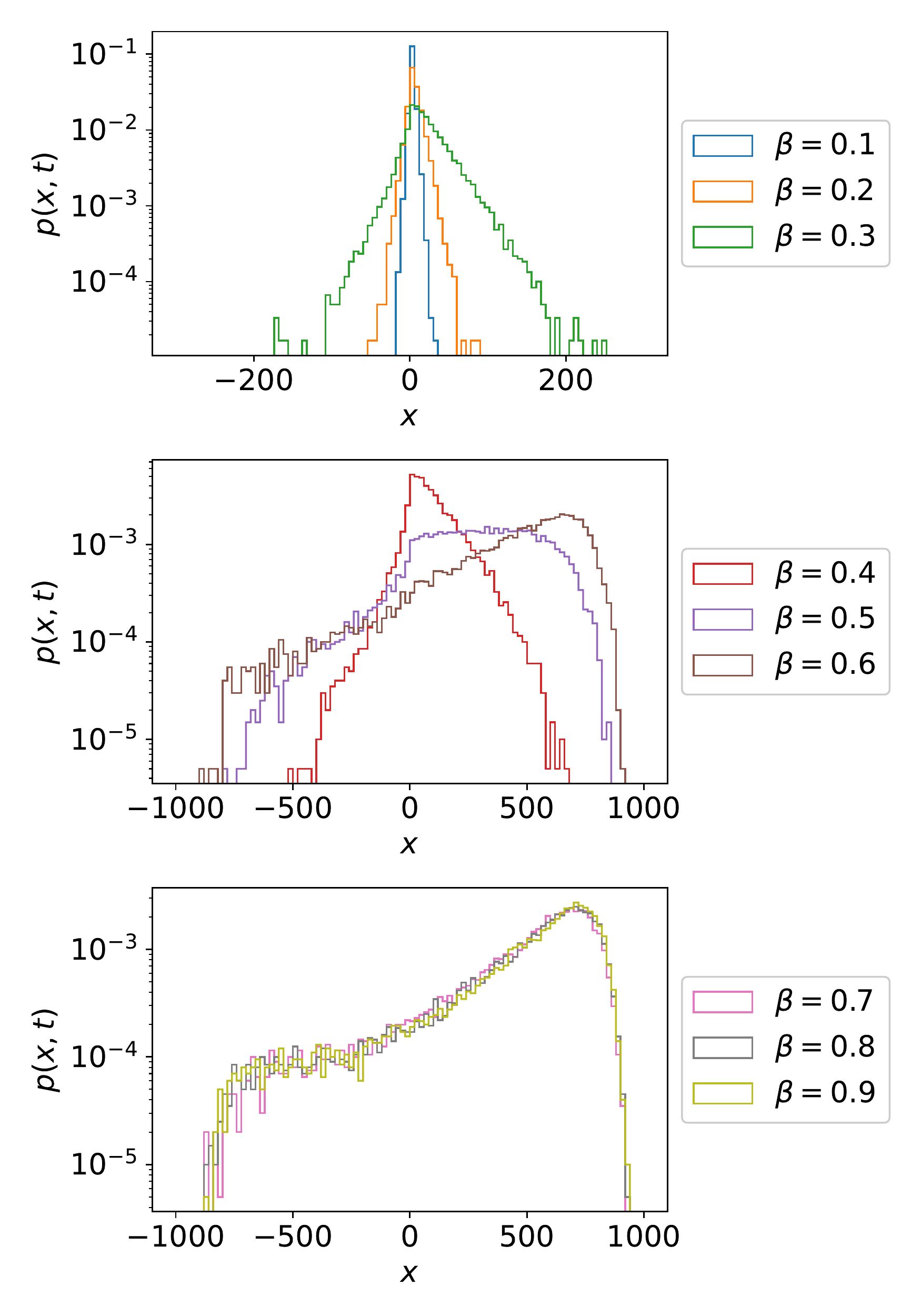} 
		\end{minipage}
		\caption{Plot of PDFs for Monte Carlo simulated trajectories at $t=10^6$ (left column) and $t=10^3$ (right column). Each value of $\beta$ has $N=10^4$ trajectories. The parameters used were $\alpha_0 = 0.6$, $\nu=1$, $\lambda =1$ and $\tau_0=1$ (left column) or $\tau_0=10^{-4}$ (right column).}
		\label{fig:4}
	\end{figure}

	This is further demonstrated by the PDFs observed in Figure \ref{fig:4}. Clearly for small values of $\tau_0$, the advection caused by self-reinforcement is dominant leading to a skewed PDF for positive velocity. However for larger $\tau_0$, the PDF reverts back to Laplacian distributions as expected for subdiffusive random walks. This finding is intriguing as self-reinforcement places a greater weight on the role of the `characteristic' scale $\tau_0$ for the power-law distributed resting times. The fact that transient superdiffusion can occur, for intermediate times in the presence of heavy tailed resting times, is suggestive of the power of this model to describe natural phenomena with time-varying anomalous exponents.
	
	
	\section{Conclusion and Summary}
	
	In this paper, we formulate a persistent random walk model for the stochastic transport of particles with self-reinforced directionality and a non-Markovian rest state with Mittag-Leffler distributed residence times. To achieve this, we derive a system of hyperbolic PDEs with a non-local switching term involving the Riemann-Liouville fractional derivative. To investigate the nature of this random walk model, we derive a fractional differential equation for the second moment. We demonstrate analytically and numerically that the introduction of anomalous rests ensures subdiffusion in the long time limit. However, transient superdiffusion is observed for intermediate times, which is also a feature in the intracellular transport of organelles. We further corroborate these results by showing the PDFs of the random walk positions, which exhibit Laplacian distributions in the long time limit but skewed bimodal distributions for intermediate times.
	
	\section{Acknowledgements}
	D.H. was funded by the Wellcome Trust grant number 215189/Z/19/Z, D.V.A was funded by the Ministry of Science and Higher Education of the Russian Federation (grant number 075-15-2021-1002). A.G. was funded by the Wellcome Trust grant number 108867/Z/15/Z. S.F. was funded by the EPSRC grant number EP/V008641/1.
	
	\bibliographystyle{unsrt}
	\bibliography{real}
	
\end{document}